\documentclass[%
reprint,
superscriptaddress,
showpacs,
amsmath,amssymb,
prl,
]{revtex4-1}
\usepackage{graphicx}
\usepackage{dcolumn}
\usepackage{bm}
\usepackage{CJKutf8}
\usepackage{color}
\usepackage{amssymb}
\usepackage{amsfonts}
\usepackage[colorlinks,urlcolor=blue,linkcolor=blue,citecolor=blue,anchorcolor=blue]{hyperref}
\makeatletter

\newcommand{\Rmnum}[1]{\expandafter\@slowromancap\romannumeral #1@}
\makeatother
\begin{document}
	
	\title{Observation of nonlinear disclination states}
	
	\author{Boquan Ren}
	\email[The two authors constribute equally to this work.]{}
	\affiliation{Key Laboratory for Physical Electronics and Devices of the Ministry of Education \& Shaanxi Key Lab of Information Photonic Technique, School of Electronic and Information Engineering, Xi'an Jiaotong University, Xi'an 710049, China}%
	
	\author{A. A. Arkhipova}
	\email[The two authors constribute equally to this work.]{}
	\affiliation{Institute of Spectroscopy, Russian Academy of Sciences, Troitsk, Moscow, 108840, Russia}%
	\affiliation{Faculty of Physics, Higher School of Economics, 105066 Moscow, Russia}%
	
	\author{Yiqi Zhang}
	\email{zhangyiqi@xjtu.edu.cn}
	\affiliation{Key Laboratory for Physical Electronics and Devices of the Ministry of Education \& Shaanxi Key Lab of Information Photonic Technique, School of Electronic and Information Engineering, Xi'an Jiaotong University, Xi'an 710049, China}%
	
	\author{Y. V. Kartashov}
	\email{yaroslav.kartashov@icfo.eu}
	\affiliation{Institute of Spectroscopy, Russian Academy of Sciences, Troitsk, Moscow, 108840, Russia}%
	
	\author{Hongguang Wang}
	\affiliation{Key Laboratory for Physical Electronics and Devices of the Ministry of Education \& Shaanxi Key Lab of Information Photonic Technique, School of Electronic and Information Engineering, Xi'an Jiaotong University, Xi'an 710049, China}%
	
	\author{S. A. Zhuravitskii}
	\affiliation{Institute of Spectroscopy, Russian Academy of Sciences, Troitsk, Moscow, 108840, Russia}%
	\affiliation{Quantum Technology Centre, Faculty of Physics, M. V. Lomonosov Moscow State University, 119991, Moscow, Russia}%
	
	\author{N. N. Skryabin}
	\affiliation{Institute of Spectroscopy, Russian Academy of Sciences, Troitsk, Moscow, 108840, Russia}%
	\affiliation{Quantum Technology Centre, Faculty of Physics, M. V. Lomonosov Moscow State University, 119991, Moscow, Russia}%
	
	\author{I. V. Dyakonov}
	\affiliation{Quantum Technology Centre, Faculty of Physics, M. V. Lomonosov Moscow State University, 119991, Moscow, Russia}%
	
	\author{A. A. Kalinkin}
	\affiliation{Institute of Spectroscopy, Russian Academy of Sciences, Troitsk, Moscow, 108840, Russia}%
	\affiliation{Quantum Technology Centre, Faculty of Physics, M. V. Lomonosov Moscow State University, 119991, Moscow, Russia}%
	
	\author{S. P. Kulik}
	\affiliation{Quantum Technology Centre, Faculty of Physics, M. V. Lomonosov Moscow State University, 119991, Moscow, Russia}%
	
	\author{V. O. Kompanets}
	\affiliation{Institute of Spectroscopy, Russian Academy of Sciences, Troitsk, Moscow, 108840, Russia}%
	
	\author{S. V. Chekalin}
	\affiliation{Institute of Spectroscopy, Russian Academy of Sciences, Troitsk, Moscow, 108840, Russia}%
	
	\author{V. N. Zadkov}
	\affiliation{Institute of Spectroscopy, Russian Academy of Sciences, Troitsk, Moscow, 108840, Russia}%
	\affiliation{Faculty of Physics, Higher School of Economics, 105066 Moscow, Russia}%
	
	\date{\today}
	
	\begin{abstract}
		\noindent Introduction of controllable deformations into periodic materials that lead to disclinations in their structure opens novel routes for construction of higher-order topological insulators hosting topological states at disclinations. Appearance of these topological states is consistent with the bulk-disclination correspondence principle, and is due to the filling anomaly that results in fractional charges to the boundary unit cells. So far, topological disclination states were observed only in the linear regime, while the interplay between nonlinearity and topology in the systems with disclinations has been never studied experimentally. We report here bon the experimental observation of the nonlinear photonic disclination states in waveguide arrays with pentagonal or heptagonal disclination cores inscribed in transparent optical medium using the fs-laser writing technique. The transition between nontopological and topological phases in such structures is controlled by the Kekul\'e distortion coefficient $r$ with topological phase hosting simultaneously disclination states at the inner disclination core and spatially separated from them corner, zero-energy, and extended edge states at the outer edge of the structure. We show that the robust nonlinear disclination states bifurcate from their linear counterparts and that location of their propagation constants in the gap and, hence, their spatial localization can be controlled by their power. Nonlinear disclination states can be efficiently excited by Gaussian input beams, but only if they are focused into the waveguides belonging to the disclination core, where such topological states reside.
	\end{abstract}
	
	\maketitle
	
	\section{Introduction}
	
	Topological systems hosting topologically protected states at their edges or in their corners are attracting considerable attention in diverse areas of physics, including solid-state physics~\cite{hasan.rmp.82.3045.2010,qi.rmp.83.1057.2011}, mechanics~\cite{huber.np.12.621.2016}, acoustics~\cite{xue.nrm.7.974.2022}, physics of matter waves~\cite{jotzu.nature.515.237.2014}, exciton-polariton condensates~\cite{klembt.nature.562.552.2018}, and, particularly, in photonics~\cite{lu.np.8.821.2014, ozawa.rmp.91.015006.2019, smirnova.apr.7.021306.2020}. This attention is connected, in part, with considerable practical potential of topological systems for construction of transmission lines, switching devices, routers, and lasers resilient to disorder and edge deformations. With the development of topological photonics~\cite{lu.np.8.821.2014, ozawa.rmp.91.015006.2019, smirnova.apr.7.021306.2020}, the class of systems, where topologically nontrivial states can be encountered has been substantially extended. While many works on photonic topological insulators employed periodic in the bulk structures for demonstration of topologically protected edge states~\cite{rechtsman.nature.496.196.2013, liang.prl.110.203904.2013, wu.prl.114.223901.2015, yang.nature.565.622.2019, maczewsky.nc.8.13756.2017, noh.prl.120.063902.2018, noh.np.12.408.2018, pyrialakos.nm.21.634.2022}, it is now realized that the topological insulators can also be created using aperiodic structures that possess discrete rotational symmetry, but lack periodicity, such as quasi-crystals~\cite{bandres.prx.6.011016.2016}, fractal structures~\cite{yang.light.9.128.2020, biesenthal.science376.eabm2842.2022}, and structures with disclinations~\cite{liu.nature.589.381.2021, peterson.nature.589.376.2021, wu.pr.9.668.2021}.
	
	The concept of topological insulators with disclinations originates from solid state physics~\cite{ruegg.prl.110.046401.2013, teo.prl.111.047006.2013, banlcazar.prb.89.224503.2014, benalcazar.prb.99.245151.2019, li.prb.101.115115.2020}, where it was predicted that disclinations --- crystallographic defects disrupting lattice structure --- can trap fractional ``spectral" charges (connected with the local density of states ~\cite{liu.nature.589.381.2021, wang.pr.9.1854.2021}) and support localized states of the topological origin. Such systems can also be used~\cite{peterson.science.368.1114.2020} for realization of the higher-order topological insulators hosting so-called zero-dimensional states~\cite{xie.nrp.3.520.2021}. The bulk-disclination correspondence principle proposed for these systems that links the appearance of the disclination states with the topological properties of the spectrum, illustrates the importance of fractional spectral charges as a probe of ``crystalline" topology of these systems~\cite{peterson.nature.589.376.2021, liu.nature.589.381.2021, chen.prl.129.154301.2022}. Higher-order topological disclination states typically form at the boundary of the hollow disclination core of the structure. It has been demonstrated that linear lattices with disclinations offer new opportunities for the control of confinement and internal structure of the field, not only in photonics~\cite{wang.prl.124.243602.2020, xie.pra.106.L021502.2022}, but also in acoustics~\cite{wang.nc.12.3654.2021, deng.prl.128.174301.2022, liang.prb.106.174112.2022}. Different from aforementioned achievements reported only in linear media, the impact of nonlinearity on photonic disclination states was addressed theoretically only recently~\cite{ren.apl.8.016101.2023}, while experimental observation of nonlinear disclination states has never been performed so far.
	
	At the same time, nonlinear effects, such as self-action of light, attract more and more attention in topological photonics~\cite{smirnova.apr.7.021306.2020} because they enable all-optical control of the properties of the topological states. New effects of topological origin emerging due to self-action include topological phase transitions~\cite{maczewsky.science.370.701.2020}, nonlinear Thouless pumping~\cite{jurgensen.nature.596.63.2021, fu.prl.128.154101.2022, fu.prl.129.183901.2022, jurgensen.np.19.420.2023}, formation of the topological solitons~\cite{ablowitz.pra.90.023813.2014, leykam.prl.117.143901.2016, ablowitz.pra.96.043868.2017, lumer.prl.111.243905.2013, zhang.nc.11.1902.2020, mukherjee.science.368.856.2020, ivanov.acs.7.735.2020, ivanov.pra.103.053507.2021, zhong.ap.3.056001.2021, ren.nano.10.3559.2021},
	development of the modulational instabilities of the edge states~\cite{kartashov.optica.3.1228.2016, zhang.pra.99.053836.2019} and rich bistability effects~\cite{kartashov.prl.119.253904.2017, zhang.lpr.13.1900198.2019}, to name only a few. Nonlinear higher-order topological insulators supporting corner solitons have been also reported theoretically~\cite{zhang.ol.45.4710.2020} and in experiment~\cite{kirsch.np.17.995.2021, hu.light.10.164.2021}, while the Floquet version of higher-order nonlinear topological insulator was proposed just recently in~\cite{zhong.pra.107.L021502.2023}.
	
	Disclination states appearing in aperiodic structures obtained by specific deformations of periodic arrays, formally belong to a class of higher-order topological states. However, in contrast to the previously considered higher-order insulator geometries with periodic bulk, disclination systems may feature other types of discrete rotational symmetries, not compatible with crystallographic symmetries and not attainable in usual higher-order insulators. One can thus expect that such symmetry properties of the disclination systems should find their manifestation in a completely different structure of their linear eigenmodes, properties of nonlinear self-sustained states bifurcating from them, and in their excitation dynamics. Our work is thus aimed at the exploration of the interplay of nonlinear self-action effects and topology in the disclination structures with different discrete rotational symmetries.
	
	Here we report on the first experimental observation of the nonlinear topological states in disclination arrays with both pentagonal and heptagonal cores, obtained by removing or adding sectors into periodic honeycomb structures, where topological phase arises due to the Kekul\'e distortion~\cite{wu.prl.114.223901.2015, noh.np.12.408.2018, liu.nature.589.381.2021, wu.pr.9.668.2021, wang.pr.9.1854.2021} introduced into positions of six sites in each unit cell of the structure. Our disclination arrays are inscribed in nonlinear fused silica, using the fs-laser direct writing technique~\cite{rechtsman.nature.496.196.2013, kirsch.np.17.995.2021, kartashov.prl.128.093901.2022, tan.ap.3.024002.2021, lin.us.2021.9783514.2021, li.ap.4.024002.2022}. We observe that when disclination arrays are in topological phase, one can excite thresholdless disclination solitons existing in a broad range of input powers by Gaussian beam focused into one of the waveguides on the disclination core. The excitation of the same waveguides in nontopological regime yields strong diffraction at low powers, while formation of nontopological self-sustained states occurs only above considerable power threshold. We thus compare behaviour of nonlinear excitations for different values of the distortion coefficient $r$. The results obtained here are relevant for a broad class of nonlinear physical systems, including matter waves, polariton condensates, photonic crystals, atomic vapors, and many others, where potentials with disclinations can be created. They also highlight the potential of these topological structures for realisation of higher-harmonic generation and lasing that may benefit from strong topological state confinement and its resilience to disorder.
	
	\begin{figure*}[htbp]
		\centering
		\includegraphics[width=1\textwidth]{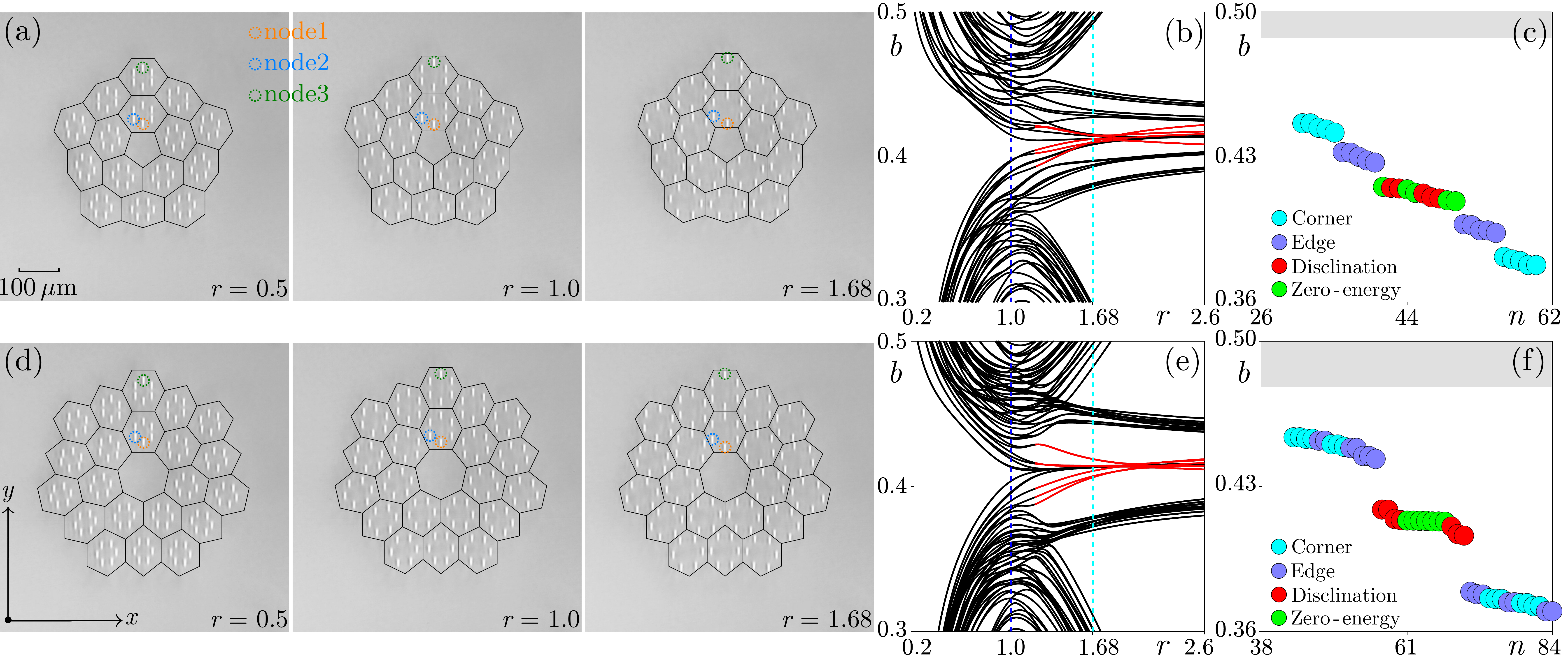}
		\caption{\textbf{Disclination arrays and their linear spectra.} 
			(a) Microphotographs of the fs-laser written waveguide arrays with a pentagonal disclination core for different values of the distortion coefficient $r$. The orange, blue, and green dotted circles indicate the nodes 1, 2 and 3, that will be used below for probing of excitation dynamics. (b) Propagation constants $b$ of the eigenmodes of pentagonal disclination array vs distortion coefficient $r$. Red curves are associated with states residing at the disclination core. (c) Spectrum at $r =1.68$. The bands corresponding to the bulk states are shown in gray, while propagation constants of corner, edge, zero-energy and disclination states are represented by dots of different colors. (d-f) Microphotographs and spectrum for the array with heptagonal disclination core. The arrangement of panels is the same as in (a)-(c).}
		\label{fig1}
	\end{figure*}
	
	\section{Results and discussions}
	\subsection{Spectra of the arrays with disclinations}
	We consider paraxial propagation of a light beam along the $z$ axis of a medium with the focusing cubic nonlinearity and shallow transverse refractive index modulation that can be described by the nonlinear Schr\"odinger-like equation for the dimensionless light field amplitude $\psi$:
	\begin{equation}\label{eq1}
		i \frac{\partial \psi}{\partial z} = -\frac{1}{2} \left( \frac{\partial^2}{\partial x^2} + \frac{\partial^2}{\partial y^2} \right) \psi
		-\mathcal{R}(x,y) \psi - |\psi|^{2} \psi,
	\end{equation}
	where $x,y$ are the scaled transverse coordinates, $z$ is the propagation distance that plays the same role as time in the Schrödinger equation describing a quantum particle in a potential, and the function ${\mathcal R}(x,y)$ describes disclination array with the straight waveguides:
	\[
	{\mathcal R} = p \sum_{m,n} e^{[-(x-x_{m,n})^2/a_x^2-(y-y_{m,n})^2/a_y^2]} ,
	\] 
	where $a_x$ and $a_y$ are the widths of waveguides that are elliptical due to the writing process, $x_{m,n}$ and $y_{m,n}$ are the positions of the waveguide centers (depending on the particular type of introduced disclination), and $p$ is the array depth proportional to the refractive index contrast $\delta n$ in the structure (see \textbf{Methods} for adopted normalizations). To create topologically nontrivial arrays with disclinations we use two-step process. We start from usual periodic honeycomb waveguide array with identical waveguide spacing $d$ in the entire structure and first introduce shift of the waveguides in the direction perpendicular to the borders of the unit cell, whose magnitude can be characterized by the Kekul\'e distortion coefficient $r = \ell_{\rm intra} / \ell_{\rm inter}$, with $\ell_{\rm intra}$ and $\ell_{\rm inter}$ being intra-cell and inter-cell spacing between waveguides after shift~\cite{wu.prl.114.223901.2015, noh.np.12.408.2018, wu.pr.9.668.2021, liu.nature.589.381.2021}. Clearly, $r=1$ corresponds to non-deformed structure with $\ell_{\rm intra} = \ell_{\rm inter} = d$. As it will be shown below, by changing the value of $r$ one can achieve the transition between topologically trivial and nontrivial geometries. On the second step, to create the arrays with disclination, we remove or add sectors into honeycomb structure with shifted waveguides. At this step, after removing of the sector we deform the unit cells in the remaining structure such that they fill the entire $2\pi$ polar angle, while to add the sector we compress unit cells accordingly (see \textbf{Methods} for the description of the deformation process).
	
	In Fig.~\ref{fig1}(a), we display the microphotographs of the arrays with pentagonal disclination core inscribed with fs-laser in $10~\textrm{cm}$ long fused silica sample (total number of waveguides in this structure is 90), obtained by removing a sector from honeycomb array, with coordinates of the waveguides obtained using the above two-step process for three different values of the distortion coefficient $r$. Black hexagons superimposed on the microphotographs are guides for the eye indicating different unit cells of the structure. Similar microphotographs, but for the structure with the heptagonal disclination core (total number of waveguides is 126) that was obtained by adding the sector into honeycomb array are presented in Fig.~\ref{fig1}(d), also for three different $r$ values. Topological properties of these structures are controlled by the distortion coefficient $r$. One can see that for $r<1$ the inter-cell coupling becomes weaker than the intra-cell one, while for $r>1$ the situation is reversed, and the inter-cell coupling becomes stronger than the intra-cell one, indicating on the possible transition of the disclination array into higher-order topological insulator phase. This transition is manifested in qualitative modification of the linear spectrum of eigenmodes supported by these structures. To obtain such modes, we first use the ansatz $\psi=u(x,y)e^{ibz}$, where $b$ is the propagation constant and $u(x,y)$ is the real function, for Eq.~(\ref{eq1}) to get the equation
	\begin{equation}\label{eq2}
		bu = \frac{1}{2} \left( \frac{\partial^2}{\partial x^2} + \frac{\partial^2}{\partial y^2} \right) u + \mathcal{R}u + u^3.
	\end{equation}
	We then omit last nonlinear term in Eq.~(\ref{eq2}) and calculate all linear eigenmodes of the system using the plane-wave expansion method.
	
	\begin{figure*}[htbp]
		\centering
		\includegraphics[width=1\textwidth]{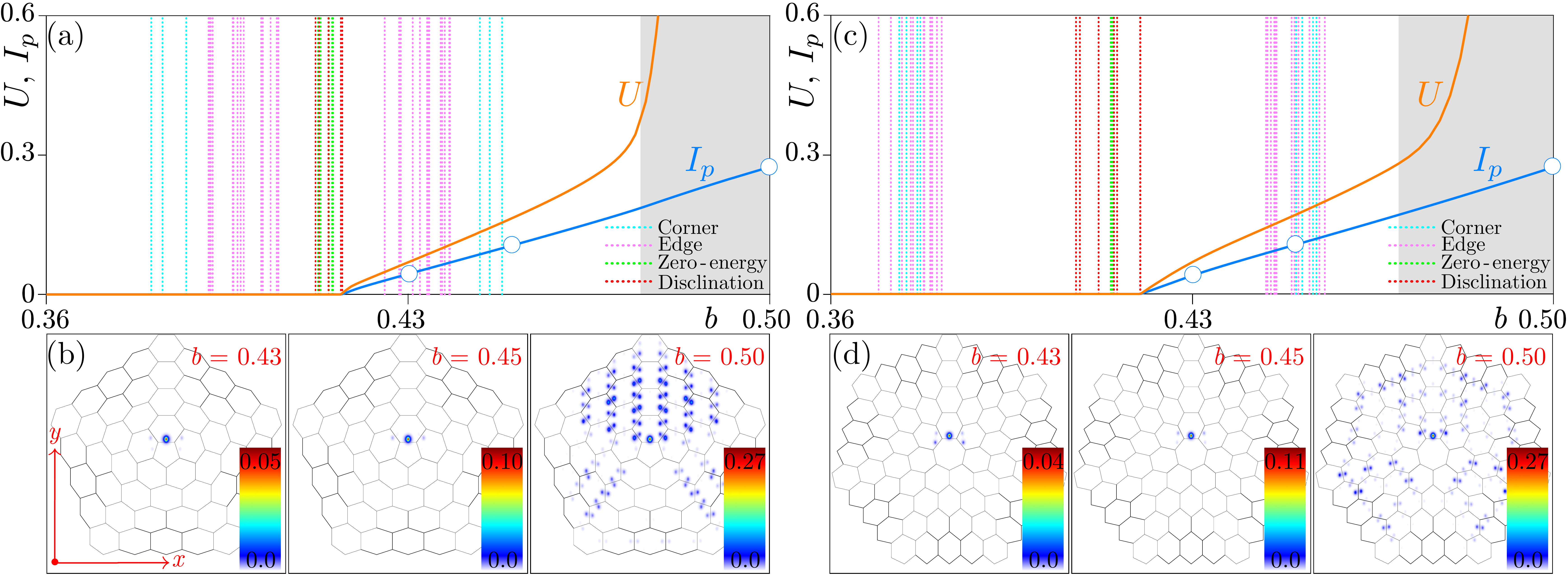}
		\caption{\textbf{The families of nonlinear disclination states.} 
			(a) Peak intensity $I_p$ (blue solid curve) and power $U$ (orange solid curve) of the nonlinear disclination states vs propagation constant $b$ in the array with the pentagonal core at $r = 1.68$. Gray regions represent the bulk band, while the vertical dotted color lines show propagation constants of linear corner, edge, zero-energy, and disclination states. (b) Intensity distributions of selected nonlinear disclination states with different propagation constants that correspond to circles in (a). (c,d) The families of nonlinear disclination states and examples of their profiles in the array with heptagonal core. Intensity distributions in (b) are shown within the window $-40\le x, y \le 40$, while those in (d) are shown within the window $-46\le x, y \le 46$.}
		\label{fig2}
	\end{figure*}
	
	The transformation of linear spectrum of the array with the pentagonal disclination core with increase of the distortion coefficient $r$ is illustrated in Fig.~\ref{fig1}(b). While at $r<1$ no localized states are present in the gap between two bulk bands, at $r>1$ the spectrum changes qualitatively, with several different types of localized states emerging in the gap. 
	The appearance of such states can be explained by the bulk-disclination correspondence principle which gives the link between the fractional charge and topologically nontrivial states (see \textbf{Methods} for discussion of the topological invariants). Among these states, \textit{five} states marked with red color (their number is dictated by the symmetry of the structure) are disclination states residing at the disclination core in the center of the array. Some of these states can be degenerate depending on the value of $r$, but in general they have different eigenvalues. These states have different phase structure, their localization at the disclination core increases with the increase of $r$. To describe the structure of the spectrum in more details, we chose  $r=1.68$ [cyan dashed line in Fig.~\ref{fig1}(b)] and show eigenvalues of all modes in the interval $0.36\le b \le 0.50$ in Fig.~\ref{fig1}(c). Besides disclination states, in the same gap there appear corner (cyan dots), edge (purple dots), and zero-energy states (green dots), but all of them emerge at the outer edge/corners of the structure due to its finite size and on this reason they do not hybridize for sufficiently large $r$ with red disclination states localized on the central disclination core. Calculated intensity distributions of all eigenmodes forming in the gap at $r=1.68$ for the arrays with pentagonal and heptagonal disclination cores are presented in \textbf{Supplemental Materials} --- to stress generality of these results, we present them for even larger structures with 300 (420) waveguides for pentagonal (heptagonal) cases --- while in experiments we use sufficiently large and most compatible with writing technology structures from Fig.~\ref{fig1}.
	
	Similar transformation of linear spectrum with increase of $r$ is observed also in the array with heptagonal disclination core, see Fig.~\ref{fig1}(d). In this structure \textit{seven} disclination states with different phase structures emerge in the spectrum (some of them are nearly degenerate so that there are seemingly five red curves) shown in Fig.~\ref{fig1}(e). Detailed structure of spectrum for this case is presented in Fig.~\ref{fig1}(f) for $r=1.68$, where one can again see that disclination states at the disclination core may coexist with spatially separated from them corner (cyan dots), edge (purple dots), and zero-energy states (green dots) at the outer edge.
	
	The emergence of disclination states of topological origin at the inner disclination core is consistent with the bulk-disclination correspondence principle~\cite{peterson.nature.589.376.2021, liu.nature.589.381.2021, chen.prl.129.154301.2022} that establishes the link between the fractional disclination charge $\mathcal{Q}$ (see \textbf{Methods} for details of topological characterization) and the localized states emerging at the disclination core. For our arrays, $\mathcal{Q}=1/2$ in topologically nontrivial phase at $r>1$ signalizing on the appearance of disclination states, while $\mathcal{Q}=0$ in nontopological regime, when $r<1$ and disclination states are absent.

	\begin{figure*}[htbp]
		\centering
		\includegraphics[width=1\textwidth]{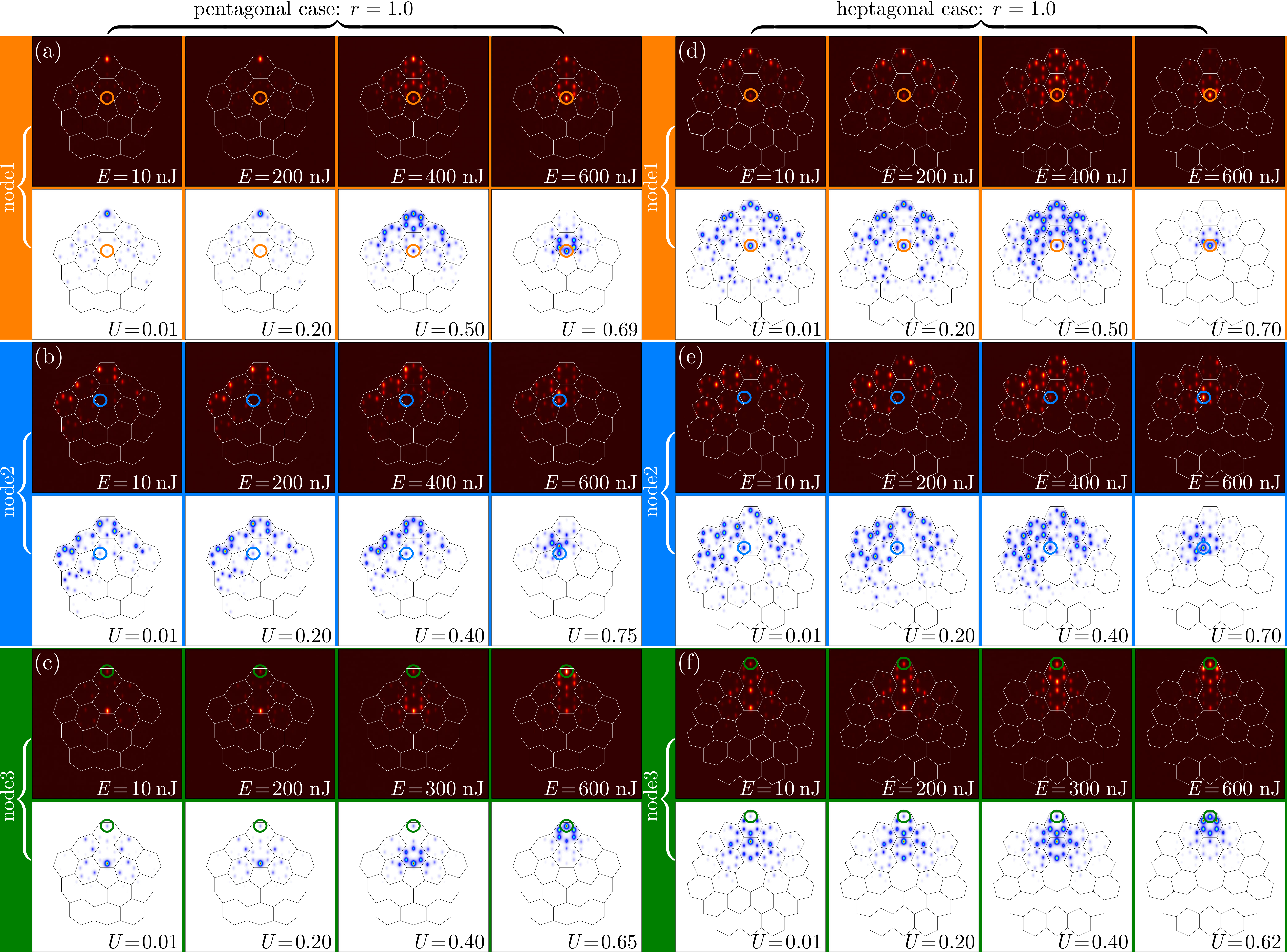}
		\caption{\textbf{Excitation of nonlinear modes in disclination arrays without distortion, at $r=1.0$.} Comparison of experimentally measured (maroon background) and theoretically calculated (white background) output intensity distributions in pentagonal (a)-(c) and heptagonal (d)-(f) disclination arrays for different input powers (pulse energies) of single-site Gaussian excitation. Results for beams focused into different nodes 1, 2 and 3 of the structure [indicated by colored circles and corresponding to the circles in Figs.~\ref{fig1}(a) and \ref{fig1}(d)] are additionally highlighted with the orange, blue, and green backgrounds. Pulse energies $E$ for the experimental outputs and dimensionless input powers $U$ for theoretical outputs are indicated on each panel. White and black lines are guides for the eye illustrating unit cells of the array. All theoretical panels are shown within the window $-30\le x,y \le 30$ and are obtained for the array depth $p=5.0$.}
		\label{fig3}
	\end{figure*}
	
	\begin{figure*}[htbp]
		\centering
		\includegraphics[width=1\textwidth]{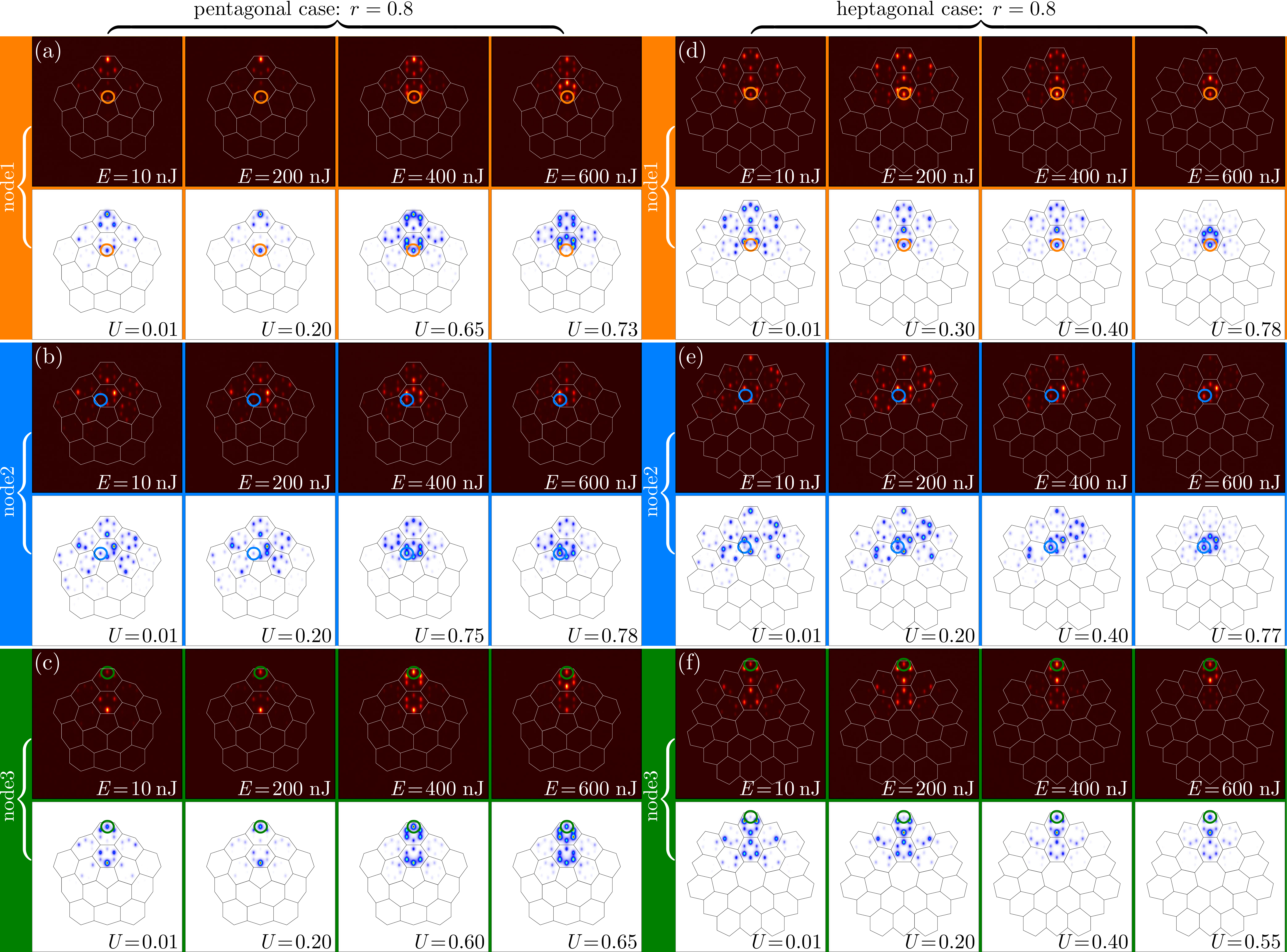}
		\caption{\textbf{Excitation of nonlinear modes in nontopological disclination array with $r=0.8$.} Comparison of the output theoretical and experimental output intensity distributions for different input powers in pentagonal (a)-(c) and heptagonal (d)-(f) arrays in topologically trivial phase. The arrangement of panels is similar to Fig.~\ref{fig3}. The array depth is $p=4.9$ in (a)-(c) and $p=5.0$ in (d)-(f).}
		\label{fig4}
	\end{figure*}
	
	\begin{figure*}[htbp]
		\centering
		\includegraphics[width=1\textwidth]{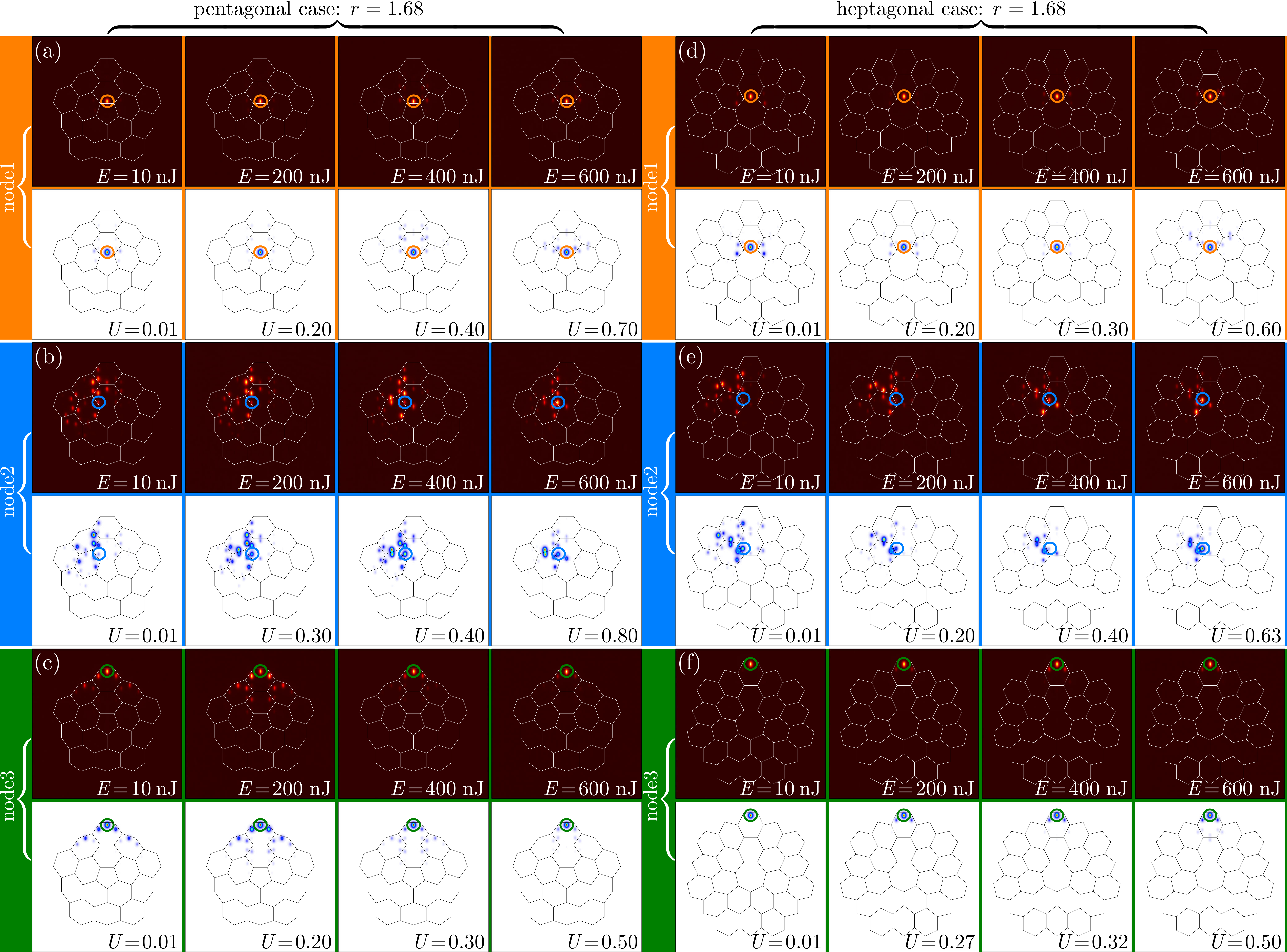}
		\caption{\textbf{Excitations of nonlinear modes in topological disclination arrays with $r=1.68$.} Comparison of experimental and theoretical outputs in pentagonal (a)-(c) and heptagonal (d)-(f) arrays in the topological phase illustrating the formation of disclination solitons (a),(d) and corner solitons (c),(f) existing for a broad range of powers, and considerable diffraction at all power levels for excitation of node 2. In all cases the array depth $p=5.0$.}
		\label{fig5}
	\end{figure*}
	
	\subsection{Properties of nonlinear disclination states}
	
	We now address the properties of stationary nonlinear disclination states, whose shapes are governed by the Eq.~(\ref{eq2}), where we keep the last nonlinear term. Such states can be found using the Newton relaxation method. By analogy with corner solitons encountered in higher-order topological insulators with periodic bulk~\cite{kirsch.np.17.995.2021, hu.light.10.164.2021}, such nonlinear disclination states can be called disclination solitons. For their theoretical description we adopt the large-scale disclination arrays schematically depicted in Fig.~\ref{fig2}. In both pentagonal and heptagonal arrays the families of the nonlinear disclination states bifurcate from linear modes localized at the disclination core. We consider bifurcation from the disclination state with the largest propagation constant (see \textbf{Supplemental Materials}). With the increase of the propagation constant $b$ the power $U=\iint |\psi|^2 dx dy$ of nonlinear disclination state monotonically grows [see Figs.~\ref{fig2}(a) and \ref{fig2}(c) for pentagonal and heptagonal cases, respectively], while the state first rapidly localizes and already at $U\sim 0.1$ concentrates practically on one side of the disclination core [see representative intensity distributions in Figs.~\ref{fig2}(b) and \ref{fig2}(d)]. This clearly stresses solitonic nature of such state, since without nonlinearity its power would be redistributed between different sites at the disclination core due to beating between several disclination states (notice that this process may be slow because eigenvalues of linear disclination states are close). The peak intensity $I_p=\max \{ |\psi|^2 \}$ of nonlinear disclination state [blue curve in Figs.~\ref{fig2}(a) and \ref{fig2}(c)] practically linearly increases with $b$. Even though propagation constant of the nonlinear disclination state crosses eigenvalues of linear edge (magenta vertical dashed lines) and corner (cyan vertical dashed lines) states, no coupling with them occurs because they are located at the outer edge of the array. However, when the propagation constant of the nonlinear disclination state penetrates into the bulk band, shown with gray color in Figs.~\ref{fig2}(a) and \ref{fig2}(c), the coupling with bulk states occurs that leads to strong expansion over entire array [see right panels in Figs.~\ref{fig2}(b) and \ref{fig2}(d)]. As a result, the power $U$ in the band rapidly increases with increase of $b$. These results illustrate that nonlinear disclination states are the modes of topological origin, whose position inside the gap and localization degree strongly depend on their power. Linear stability analysis performed for these nonlinear families shows that they are stable in the entire gap, in both pentagonal and heptagonal structures, thus they can be readily excited in the experiment.
	
	\subsection{Observation of nonlinear disclination states}
	
	For observation of nonlinear disclination states we inscribed (see \textbf{Methods} for details of fabrication) the arrays with pentagonal and heptagonal disclination cores with different values of the distortion coefficient $r=0.8$, $1.0$ and $1.68$, to be able to compare dynamics in topologically trivial and nontrivial structures. In experiments, we employed single-waveguide excitations using $280\,\textrm{fs}$ pulses of variable energy $E$ from $1\,\textrm{kHz}$ fs Ti:sapphire laser at $800\,\textrm{nm}$ central wavelength. The input peak power in the waveguide defined as a ratio of the pulse energy $E$ to the pulse duration $\tau$ taking into account the losses for matching with the focusing lens is evaluated as $2.5\,\textrm{kW}$ for each $1\,\textrm{nJ}$. We compare excitations of three different waveguides (nodes) numbered 1 (at the disclination core), 2 (in the bulk), and 3 (in the outer corner) indicated by colored circles in Figs.~\ref{fig1}(a) and \ref{fig1}(d).
	
	In Fig.~\ref{fig3} we present comparison of the output intensity distributions for these three types of excitations calculated theoretically (images with white background) using Eq.~(\ref{eq1}) and measured experimentally (images with maroon background) for both pentagonal and heptagonal disclination arrays without distortion, i.e. with $r=1$. In this ``borderline" case between topological and nontopological phases, no localized states are present in spectra of the arrays. On this reason the excitation of any of the nodes 1, 2, or 3 is accompanied by strong diffraction in the linear regime for $E=10\, \rm nJ$ pulses (in Fig.~\ref{fig3} the position of excitation in each case is marked by colored ring). Increasing pulse energy (power in theoretical simulations) results in gradual contraction of light towards the excited waveguide. For excitation of node 1 at disclination core one can observe the formation of well-localized soliton at highest shown pulse energy $E\sim 600\, \rm nJ$, i.e. above considerable threshold [Figs.~\ref{fig3}(a) and \ref{fig3}(d)]. The same pulse energy level in general is not sufficient for soliton formation for excitation in the bulk [Figs.~\ref{fig3}(b) and \ref{fig3}(e)] and at the outer corner [Figs.~\ref{fig3}(c) and \ref{fig3}(f)], since at this energy level the tendency for light contraction to the excited waveguide only begins. To achieve good localization in these cases one has to increase the pulse energy even further, approximately to $E\sim 900\, \rm nJ$ level.
	
	In Fig.~\ref{fig4} we consider the same three types of excitations in the trivial insulator phase, when distortion coefficient $r=0.8$. According to the Wannier center analysis in each unit cell, the filling anomaly does not occur in this case, and, consequently, no localized corner, edge or disclination states can appear in the linear spectrum of the system, despite the fact that forbidden gap opens for this value of $r$ [see Figs.~\ref{fig1}(b) and \ref{fig1}(e)], i.e. all linear eigenmodes are delocalized bulk modes. Thus, one again observes diffraction in the linear regime for $E=10\, \rm nJ$ pulses, for both pentagonal Figs.~\ref{fig4}(a)-\ref{fig4}(c) and heptagonal Figs.~\ref{fig4}(d)-\ref{fig4}(f) disclination arrays for all three types of excitations. Moreover, now localization does not occur for excitation at the disclination core even for pulse energies $E\sim 600\, \rm nJ$ that was sufficient for nonlinear localization at $r=1.0$. Thus, there exists the tendency for increase of the pulse energy required for localization at the disclination core with decrease of $r$. For depicted pulse energies localization was not observed neither for bulk nor for corner excitations (it occurs only around $E\sim 900\, \rm nJ$).
	
	The picture changes qualitatively at $r=1.68$ in topologically nontrivial phase. In this case, the disclination core supports topologically nontrivial localized disclination states, thus the input beam focused into node 1 excites localized states even at the lowest pulse energies $E \sim 10\, \rm nJ$ in both pentagonal [Fig.~\ref{fig5}(a)] and heptagonal [Fig.~\ref{fig5}(d)] arrays. Notice that even though in this quasi-linear regime single-site excitation leads to simultaneous population of several localized disclination eigenmodes, the beating between them occurs on the scale much larger than sample length (due to small difference of propagation constants $b$ of such eigenmodes, see \textbf{Supplementary Materials}) and is therefore not visible in experiment at $10~\textrm{cm}$ of propagation. Even weak nonlinearity suppresses this beating leading to the formation of well-localized disclination solitons that exist over broad range of input pulse energies (powers), as long as propagation constants of such states remain in the forbidden gap of the spectrum [Figs.~\ref{fig5}(a) and \ref{fig5}(d)]. Notice that because for $r=1.68$ the gap is already wide, in experiment we do not reach power levels (below optical damage threshold), at which strong coupling with bulk states occurs. In contrast, when node 2 in the bulk is excited, one observes diffraction, and nonlinear localization does occur even for pulse energies $E\sim 600\, \rm nJ$ [see Figs.~\ref{fig5}(b) and \ref{fig5}(e)]. An interesting situation is encountered for excitation of the node 3 [Figs.~\ref{fig5}(c) and \ref{fig5}(f)]. This excitation has the largest overlap with the corner states that are also well-localized for this value of $r$ at the outer edge of the array, and it does not excite zero-energy states (because the latter have different symmetry, see \textbf{Supplementary Materials}). As a result, in this case one observes the formation of nonlinear corner states in both pentagonal and heptagonal disclination arrays, whose localization degree only weakly changes in the considered range of input powers. Theoretical simulations fully support these observations.
	
	\section{Conclusions}
	In conclusion, we have reported on the experimental observation of nonlinear disclination states in disclination lattices inscribed in transparent nonlinear optical medium. Such states form when Kekul\'e distortion of waveguide positions drives the array into the topological phase, where several different types of localized states appear: disclination states residing at the disclination core and not overlapping with them spatially corner, edge, and zero-energy states at the outer edge of the structure. The nonlinearity enables strong light localization on one side of the disclination core. Our findings are reported for the pentagonal and heptagonal structures, with symmetry different from previously considered higher-order insulators with periodic bulk (such as $\mathcal{C}_3$, $\mathcal{C}_4$, $\mathcal{C}_6$ ones). They pave the way for the development of new types of topological lasers on disclination states, efficient harmonic generation in topologically protected states, and observation of new interesting topological objects, such as topological disclination bound states in the continuum~\cite{qin.arxiv.2023}.
	
	\section{Materials and methods}
	
	\subsection{Fs-laser inscription of the waveguide arrays}
	
	The waveguide arrays shown in Figs.~\ref{fig1}(a,d) were inscribed in $10\,\textrm{cm}$-long fused silica glass samples (JGS1) using focused (with an aspheric lens with $\textrm{NA}=0.3$) under the surface of sample at the depth range of $550\sim1050\,\mu\textrm{m}$ fs laser pulses at the wavelength of $515\,\textrm{nm}$ with the duration $280\,\textrm{fs}$, repetition rate $1\,\textrm{MHz}$, and energy $320\,\textrm{nJ}$. Translation of the sample during the writing process of each waveguide was performed by the high-precision air-bearing positioner (Aerotech) with identical for all waveguides velocity of $1\,\textrm{mm/s}$. All such waveguides are elliptical and single-mode, and they exhibit propagation losses not exceeding $0.3\,\textrm{dB/cm}$ at $\lambda=800\,\textrm{nm}$. After the waveguide arrays were inscribed, the input/output facets of the sample were optically polished, so that the sample length was shortened to $99\,\rm mm$.
	
	\subsection{Numerical simulations and normalizations}
	
	For numerical simulations of evolution and excitation of the nonlinear disclination states, we used dimensionless continuous nonlinear Schr\"odinger-like equation (\ref{eq1}), in which the transverse coordinates $x,y$ are normalized to the characteristic scale $r_0=10\,\mu\textrm{m}$, the propagation distance $z$ is normalized to the diffraction length $kr_0^2\approx1.14\,\textrm{mm}$, $k=2\pi n/\lambda$ is the wavenumber in the medium with the background refractive index $n$ (for fused silica $n\approx 1.45$), and $\lambda=800~\textrm{nm}$ is the working wavelength. Our waveguides are single-mode and elliptical due to the writing process, the dimensionless widths of the waveguides are $a_x=0.25$ and $a_y=0.75$ (corresponding to $2.5$ and $7.5\,\mu\textrm{m}$, respectively), but the eigenmode of such waveguides is only slightly elliptical. The waveguide spacing in structure without distortion is $d=3.2$ (corresponding to $32\,\mu\rm m$). The array depth $p=k^2r_0^2\delta n/n$ is proportional to the refractive index contrast $\delta n$ in the structure. For instance, $p=1.0$ corresponds to $\delta n\sim 1.1\times10^{-4}$. In the majority of presented results (unless specifically stated in the caption), we use the depth $p=5.0$ that provides the best agreement between experiments and theory.
	
	\begin{figure}[htbp]
		\centering
		\includegraphics[width=1\columnwidth]{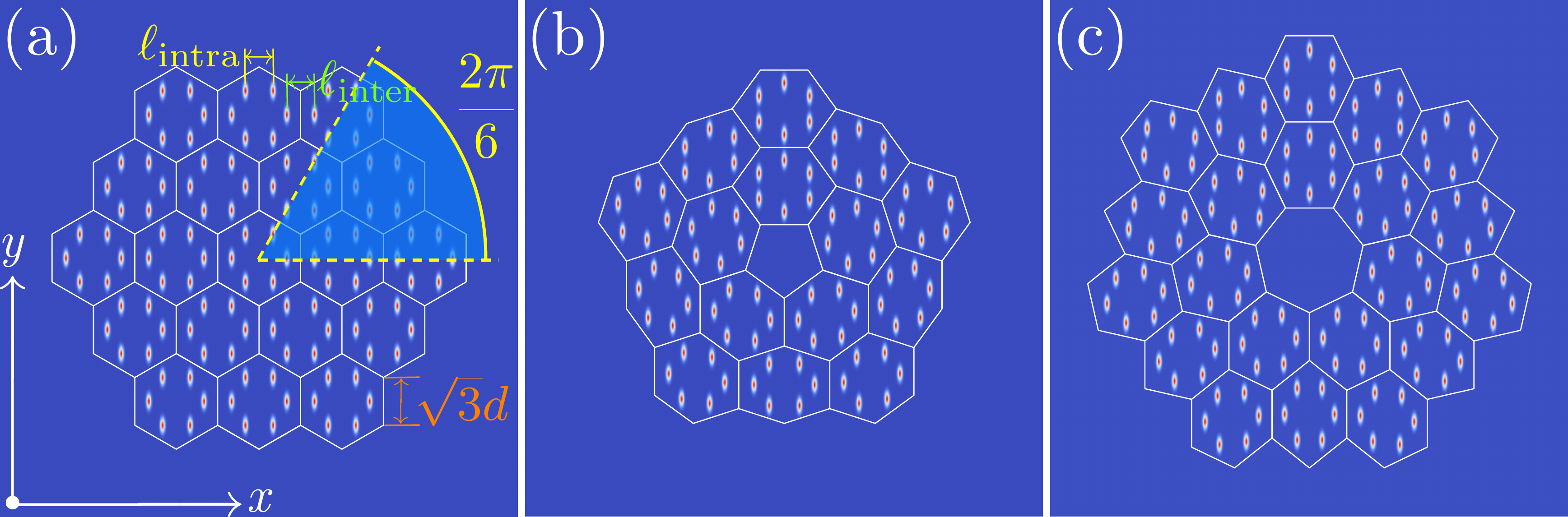}
		\caption{\textbf{Construction of the disclination array from honeycomb structure.} (a) Original honeycomb array with intra-cell separation $\ell_{\rm intra}$ and inter-cell separation $\ell_{\rm inter}$. The Kekul\'e distortion coefficient $r=\ell_{\rm intra}/\ell_{\rm inter}$. Each unit cell is indicated by white hexagons. $\sqrt{3}d$ denotes the length of one side of the hexagonal unit cell. (b) Disclination array with the pentagonal core obtained by removing the sector with the Frank angle $2\pi/6$ from array in (a) and gluing the cutting edges. (c) Disclination array with the heptagonal core obtained by inserting the sector with the Frank angle $2\pi/6$ into array (a) and subsequent compression of the unit cells. All arrays are shown within the window $-30\le x, y \le 30$.}
		\label{fig6}
	\end{figure}
	
	Pentagonal and heptagonal disclination arrays were obtained from regular honeycomb arrays using two-step process described in the main text, when at the first stage by shifting the waveguides one introduces controllable Kekul\'e distortion, quantified by the distortion coefficient $r = \ell_{\rm intra} / \ell_{\rm inter}$, where $\ell_{\rm intra}$ and $\ell_{\rm inter}$ is the intra-cell and inter-cell spacing between waveguides after shift [see notations in Fig.~\ref{fig6}(a)], while at the second stage one removes or inserts Frank sector into the array, and then expands or contracts unit cells such as to obtain the resulting disclination structure [Fig.~\ref{fig6}(b) and \ref{fig6}(c)]. While deforming the structure we keep the longer axes of all elliptical waveguides parallel to the $y$ axis.
	
	\subsection{Topological indices}
	
	The topological properties of disclination arrays can be discussed by analyzing the fractional ``charge'' that is carried by each unit cell, that is employed in established bulk-disclination correspondence principle~\cite{liu.nature.589.381.2021}. Note that the ``charge'' here is spectral charge that can be defined through the local density of states. It is an analog of the real charge in electric systems, and it can be used to evaluate the number of states in the unit cell with states considered below the topological band gap~\cite{wang.pr.9.1854.2021}.
	The spectral charge  $\mathcal{Q}$ bound to a disclination with a Frank angle
	$\Omega$ is defined by~\cite{noh.np.12.408.2018, benalcazar.prb.99.245151.2019, li.prb.101.115115.2020, peterson.nature.589.376.2021, liu.nature.589.381.2021, wu.pr.9.668.2021}
	\begin{equation}
		\mathcal{Q} = \frac{\Omega}{2\pi} \left(\frac{3}{2}\chi_{\rm M}-\chi_{\rm K} \right) \, {\rm modulo}\, 1,
	\end{equation}
	where the high symmetry indicators are $\chi_{\rm M}=\#{\rm M}_1^{(2)}-\#\Gamma_1^{(2)}$ and $\chi_{\rm K}=\#{\rm K}_1^{(3)}-\#\Gamma_1^{(3)}$ that should be calculated directly in the honeycomb array before removing or inserting the Frank sector. Here $\#\Pi_q^{(n)}$ is the number of bands below the forbidden gap at a high-symmetry point $\Pi = \Gamma, \,\rm M,\, K$ with the eigenvalue of the $C_n$ rotation matrix $e^{i2\pi(q-1)/n}~(q=1,\cdots,n)$~\cite{benalcazar.prb.99.245151.2019},
	and $\Omega=2\pi/6$ for the disclination arrays adopted in this work. 
	For the topological nontrivial case with $r>1$, one 
	can find that $(\chi_{\rm M}, \chi_{\rm K})=(2,\,0)$ for both the pentagonal disclination array and the heptagonal disclination array.
	While for the topological trivial case with $r<1$, $(\chi_{\rm M}, \chi_{\rm K})=(0,\,0)$.
	Thus, the fractional charge is $\mathcal Q=1/2$ for $r>1$ and $\mathcal Q=0$ for $r<1$. 
	The fractional charge can be also obtained by counting the number of the Wannier centers occupied by each unit cell. The Wannier centers are located at the edges of the unit cell if $r>1$ and at its center if $r<1$. The unit cell around the pentagonal disclination core has five bulk Wannier centers if $r>1$, which give a $5/2$ charge per unit cell. If $r<1$, the charge per unit cell around the disclination core is $3$. See also the \textbf{Supplementary Materials}.
	
	\section*{Acknowledgements}
	This research was funded by the National Natural Science Foundation of China (Grant No.: 12074308), the research project FFUU-2021-0003 of the Institute of Spectroscopy of the Russian Academy of Sciences and partially by the RSF grant 21-12-00096, the Foundation for the Advancement of Theoretical Physics and Mathematics “BASIS” (Grant No.: 22-2-2-26-1),	and the Fundamental Research Funds for the Central Universities (Grant No.: xzy022022058).
	
	\section*{Author contributions}
	All authors made significant contribution to this work.
	
	\section*{Conflict of interest}
	The authors declare no competing interests.
	
	\section*{Supplementary information} 
	The online version contains supplementary
	material available at https://doi.org/.
	
	\bibliography{my_library}
	\bibliographystyle{myprl}

\end{document}